\documentstyle[twocolumn,prl,aps]{revtex}
\begin{document}
\draft
\title 
{Temperature dependent correlation length for the S=1/2 Quantum
 Heisenberg Antiferromagnet on the square lattice}

In a recent Letter\cite{CUCC}, Cuccoli {\it et al} presented a new
theoretical approach to the understanding of the two dimensional (2D)
quantum Heisenberg antiferromagnet (QHA). 
The approach is characterized by the feasibility of the
separation of the quantum fluctuation from the thermal one.
One of their main predictions is that $\xi$ for the 
QHA is given by that of the classical counterpart, i.e.,
\begin{equation}
\xi(t) = \xi_{cl}(t_{cl}),~~~t_{cl}=t/\theta^{4}(t), \label{eq:fun}
\end{equation}
where $t=kT/J$ for $S=1/2$ and $\theta(t)$ can be determined
from their theory using their Eqs.(6)-(8), at least numerically. 

The authors concluded that their theory ``appears to explain
all the experimental data for different values of $S$ 
without any fitting parameters''. 
Consequently, they questioned the validity of the key assumption 
of the conventional theoretical approach\cite{CHA}. 
This conclusion, however, 
is surprising because the Gaussian approximation they used to 
handle the quantum fluctuation is supposed to be accurate 
only when the effect of the fluctuation is weak, i.e., 
for sufficiently large value of  $S$ and high temperature.
Since the experimental data have rather large 
statistical fluctuation
and previously available MC data limited to $\xi < 30$ are not
completely free of systematic errors,
much more accurate data including larger $\xi$
appear to be crucial to check the validity of their theory.

In this Comment we present high precision MC data of the second moment
correlation length for $S=1/2$ QHA on $L \times L$ lattices, up to
$\xi = 95.7(3)$.  Our data are obtained using a powerful new quantum
MC method\cite{BEARD}
which completely eliminates the systematic error coming 
from finite Suzuki-Trotter number. 
We carefully monitored the finite size effect in our data by repeating
measurements on varying lattice size from L= 20 to 1000 and found that
it becomes smaller than the typical statistical error of 0.3 percent
or better under the condition $L/\xi \gtrsim 7$.
%Part of our thermodynamic (infinite volume limit)  data 
%are tabulated in Table(1).

The classical correlation length $\xi_{cl}(t_{cl})$ can be determined
from extensive MC data already available for the 2D classical
Heisenberg ferromagnet\cite{APO} much more precisely than using the
data given by Cuccoli {\it et al.}\cite{CUCC}, because an extrapolation
of $\xi_{cl}(t_{cl})$ for smaller $t_{cl}$ is necessary. Considering
$\xi_{cl}$ over the range $34 \lesssim \xi_{cl} \lesssim
788$\cite{APO} (corresponding to $0.20 \lesssim t \lesssim 0.29$), we
find a good fit assuming the functional form from the two-loop order
of the perturbation theory for the classical model, i.e.,
$\xi_{cl}(t_{cl}) = C_{\xi}~t_{cl}~e^{a/t_{cl}}
\left[1 + c_{1}~t_{cl} + \ldots \right] \label{eq:clcor}$,
with $C_{\xi}\simeq 2.300\times 10^{-3}$, $a\simeq 6.378$, $c_{1}
\simeq -0.855$, and $\chi^{2}/N_{DF}$ (the $\chi^{2}$ value per degree
of freedom) $\simeq 0.4$.  For the $\xi_{cl}$ over $2 \lesssim
\xi_{cl} \lesssim 34$ (corresponding to $ 0.29 \lesssim t \lesssim
0.65$), on the other hand, it turns out that an additional correction
of the type $c_{2}t_{cl}^{2}$ is
necessary for a good fit.  The estimated fitting parameters in this
case are: $C_{\xi}\simeq 8.129 \times 10^{-3}$, $a\simeq 6.050$,
$c_{1} \simeq -2.862$, and $c_{2} \simeq 2.393$.
Given $\xi_{cl}(t_{cl})$, we obatin $\xi(t)$ directly from Eq.(1) for
all the values of $t$ where our quantum Monte Carlo data are
available.

In Fig. (1) we present our results. 
We see a good agreement with the
theory for $\xi(t) \lesssim 6$. However, our data clearly deviate
from the theory for larger values of $\xi$. In fact the theory
overestimates the actual numerical data more than 500 percent at
$t\simeq 0.21$. We thus conclude that the validity of 
the new approach is limited to very high temperature regime,
i.e., $T/J \gtrsim 0.4$ for the 2D $S=1/2$ QHA.

We would like to thank A. Cuccoli, V. Tognetti, R. Vaia and
P. Verrucchi for helpful comments and for sending us their numerical
estimates for the function $\theta^{4}$.

%\begin{table}
%\caption{Thermodynamic data of $\xi$ for the 2D $S=1/2$ QHA, over $3.25
% \le J/T  \le 4.50$.}
%\begin{tabular}{ccccccc}
%J/T &3.25  &3.50 &3.75 &4.00 &4.25  &4.50 \\
%$\xi$ &15.7(1) &21.4(1) &29.0(1) &39.2(1) &52.8(2) &70.9(2) 
%\end{tabular}
%\end{table} 
\par
\medskip\noindent
Jae-Kwon Kim$^a$, D. P. Landau$^a$, and Matthias Troyer$^b$ \par
\vspace{0.2cm}
$^a$Center for Simulational Physics \par
The University of Georgia,  Athens, GA 30602 \par
\vspace{0.2cm}
$^b$Institute for Solid State Physics\par
University of Tokyo, Tokyo 106, Japan

%\begin{figure}
%\caption{
\vspace{0.3cm}
{\bf Figure Caption:} Comparison of the Monte Carlo data with the theoretical 
         prediction Eq.(1) for the 2D $S=1/2$ QHA.
         The statistical errors of our data are much smaller
         than the size of the symbol.
%\label{fig:Cucc_out.eps}
%\end{figure}

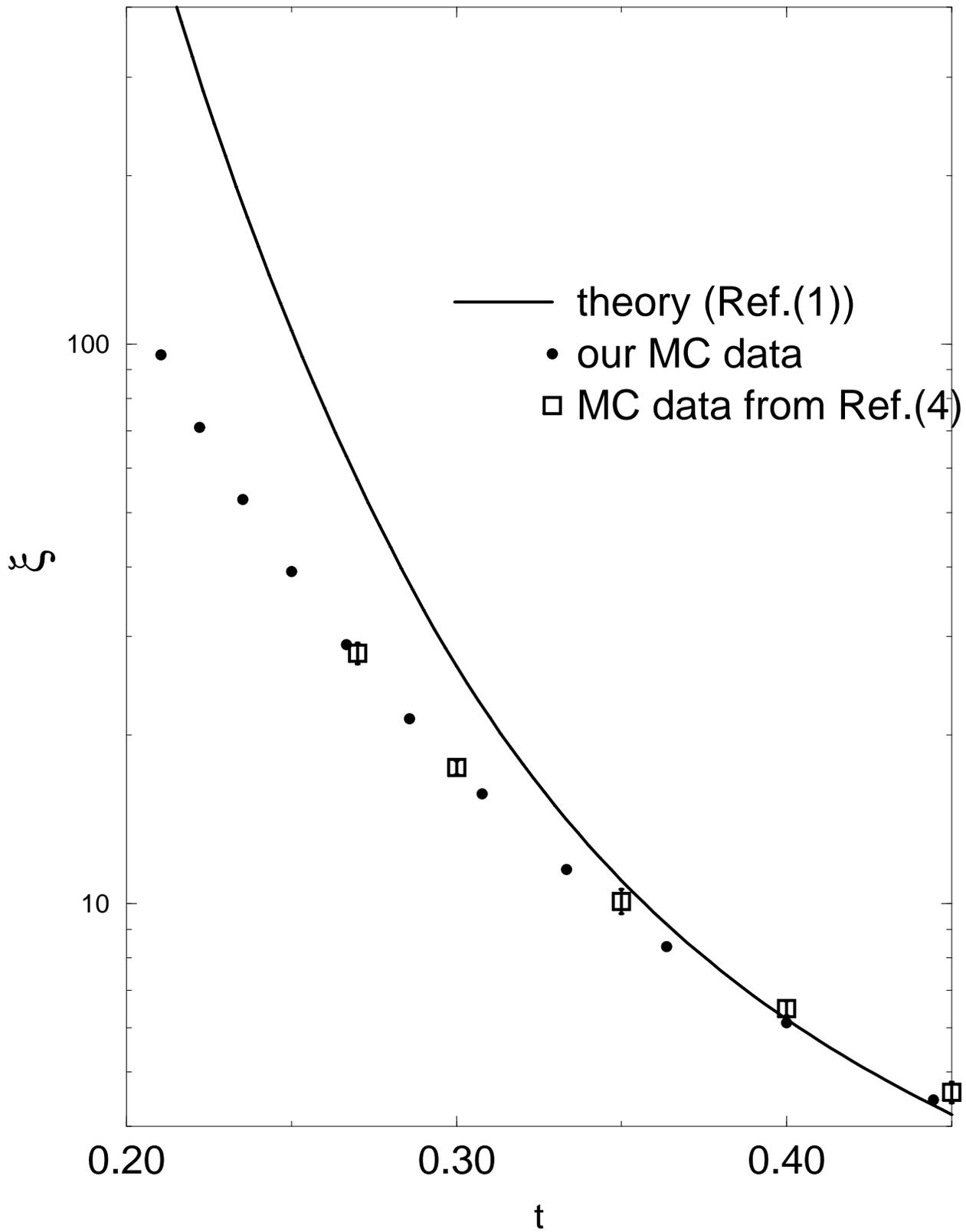
\end{document}